\documentclass[aps, reprint, superscriptaddress, longbibliography]{revtex4-1}

\usepackage{amsmath}
\usepackage{amssymb}
\usepackage{siunitx}
\usepackage{placeins}
\usepackage{physics}
\usepackage{graphicx}
\usepackage{multirow}
\usepackage{hyperref}
\usepackage{xcolor}

\DeclareMathOperator*{\vari}{var}
\DeclareMathOperator*{\cov}{cov}

\begin{document}
\title{All rf-based tuning algorithm for quantum devices using machine learning}

\author{Barnaby van Straaten}
\thanks{These authors contributed equally.}
\affiliation{Department of Materials, University of Oxford, Oxford OX1 3PH, United Kingdom}

\author{Federico Fedele}
\thanks{These authors contributed equally.}
\affiliation{Department of Materials, University of Oxford, Oxford OX1 3PH, United Kingdom}

\author{Florian Vigneau}
\affiliation{Department of Materials, University of Oxford, Oxford OX1 3PH, United Kingdom}

\author{Joseph Hickie}
\affiliation{Department of Materials, University of Oxford, Oxford OX1 3PH, United Kingdom}

\author{Daniel Jirovec}
\affiliation{Institute of Science and Technology Austria, Am Campus 1, 3400 Klosterneuburg, Austria}

\author{Andrea Ballabio}
\affiliation{L-NESS, Physics Department, Politecnico di Milano, via Anzani 42, 22100, Como, Italy}

\author{Daniel Chrastina}
\affiliation{L-NESS, Physics Department, Politecnico di Milano, via Anzani 42, 22100, Como, Italy}

\author{Giovanni Isella}
\affiliation{L-NESS, Physics Department, Politecnico di Milano, via Anzani 42, 22100, Como, Italy}

\author{Georgios Katsaros}
\affiliation{Institute of Science and Technology Austria, Am Campus 1, 3400 Klosterneuburg, Austria}

\author{Natalia Ares}
\email{natalia.ares@eng.ox.ac.uk}
\affiliation{Department of Engineering Science, University of Oxford, Oxford OX1 3PJ, United Kingdom}

\date{\today}

\begin{abstract}

Radio-frequency measurements could satisfy DiVincenzo's readout criterion in future large-scale solid-state quantum processors, as they allow for high bandwidths and frequency multiplexing. However, the scalability potential of this readout technique can only be leveraged if quantum device tuning is performed using exclusively radio-frequency measurements i.e. without resorting to current measurements. We demonstrate an algorithm that automatically tunes double quantum dots using only radio-frequency reflectometry. Exploiting the high bandwidth of radio-frequency measurements, the tuning was completed within a few minutes without prior knowledge about the device architecture. Our results show that it is possible to eliminate the need for transport measurements for quantum dot tuning, paving the way for more scalable device architectures.

\end{abstract}

\maketitle
\section{Introduction}

Radio-frequency (rf) reflectometry allows for high-bandwidth measurements of quantum devices \cite{vigneau_probing_2022}. Due to their potential for scalability, rf techniques play an increasingly important role in developing quantum device circuits. A recent achievement has been the high-fidelity single-shot readout of spins \cite{hogg_single-shot_2022, madzik_precision_2022}. However, until now, the much slower measurement of current through the device was still necessary to automatically tune quantum devices to their operating condition \cite{moon_machine_2020, baart_computer-automated_2016, mills_computer-automated_2019, zwolak_autotuning_2020}. This was a limiting factor for scalability because such measurements are incompatible with scalable device architectures \cite{cai_silicon_2019, li_crossbar_2018, veldhorst_silicon_2017, gonzalez-zalba_scaling_2021} and are too slow for error correction \cite{laucht_roadmap_2021}. 

We demonstrate an algorithm that tunes a quantum device using rf measurements exclusively, eliminating the need for current measurements through the device. This opens up the possibility of new, more scalable device architectures designed only to use rf measurements. The algorithm aimed to define double quantum dots, a cornerstone of the solid-state qubit effort. These devices can encode either a singlet-triple qubit or a pair of Loss-Divincenzo spin qubits.  \cite{chatterjee_semiconductor_2021, burkard_semiconductor_2021}.

Our algorithm exploits the bandwidth of rf measurements to acquire high-resolution two-dimensional charge stability diagrams within milliseconds \cite{stehlik_fast_2015, schupp_sensitive_2020}. The device's gate voltage space is efficiently explored using Gaussian processes, principal component analysis to perform blind signal separation \cite{jolliffe2016principal, madhab2013, comon_2010}, and Kolmogorov-Smirnov statistical tests \cite{hodges_significance_1958}, along with our fast scans. Quantum dot features are identified using a score function based on the Fourier transform of the charge stability diagram. 

We demonstrate our tuning algorithm in a hole-based quantum dot array hosted in a Ge/SiGe heterostructure. Ge hole spin qubits are currently enjoying intense research interest owing to their ease of operation and compatibility with existing Si technology \cite{scappucci_germanium_2021}. From 2018, and within just four years, a Loss-DiVincenzo qubit \cite{watzinger_germanium_2018}, a single-triplet qubit \cite{jirovec_singlet-triplet_2021}, two-qubit devices \cite{hendrickx_fast_2020} and a four-qubit Ge quantum processor \cite{hendrickx_four-qubit_2021} have been realised, demonstrating the potential of Ge for quantum information. Furthermore, Ge has also hosted record-breaking ultra-fast control frequencies, with Rabi frequencies exceeding \SI{540}{\mega \hertz} \cite{wang_ultrafast_2022}. 

Our algorithm contributes significantly towards scalable readout, an essential step in realising the scalability potential of Si-compatible quantum circuits.

\begin{figure*}
    \centering
    \includegraphics{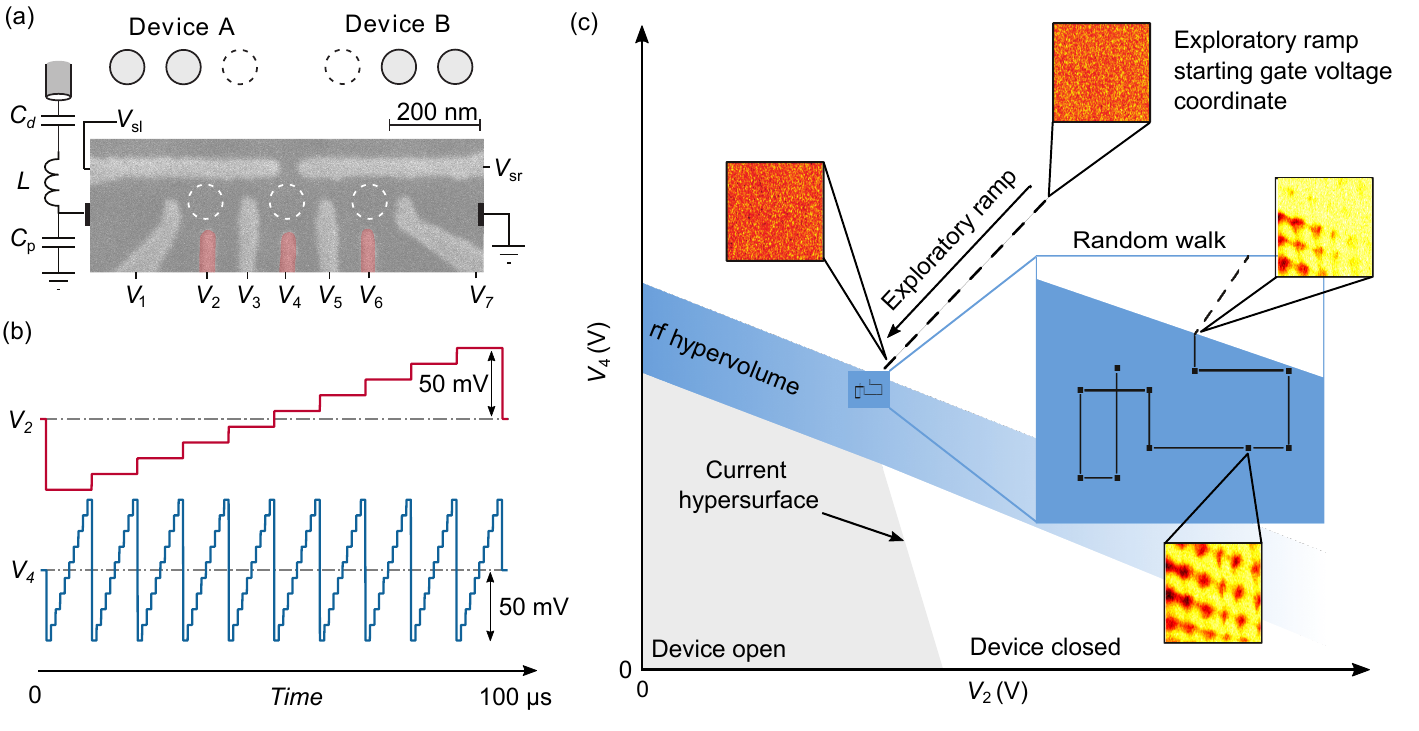}
    \caption{(a) A scanning electron micrograph of the Ge/SiGe quantum dot array, with an L-matching circuit connected to the device ohmic. Bias-tees are used on gates shaded in red to apply both ac and dc voltages. Devices A and B demonstrate how two different double dots can be defined in the array. (b) A diagram showing the waveforms sent to gates $V_2$ ($V_4$) and $V_4$ ($V_6$) to perform a fast two-dimensional (2D) scan of $10 \times 10$ pixels and $100\times100$ \SI{}{\milli \volt} on device A (B). (c) An illustrative diagram showing an early iteration of the algorithm. For clarity, we restrict the schematic to the plane defined by the two fast plunger gates of a device. Firstly, the algorithm performs an exploratory ramp along a randomly chosen direction, where a Gaussian process chooses the starting gate voltage coordinate to lie as close to, but still outside of, the rf hypervolume as its confidence allows. Along the length of the exploratory ramp, fast 2D scans are performed using gates $V_2$ ($V_4$) and $V_4$ ($V_6$) for device A (B). The rf classifier evaluates each of these scans to determine whether they show rf features or just noise. The first instance of the rf classifier classifying features marks the point where the exploratory ramp enters the rf hypervolume and the rf becomes sensitive. At this point, the exploratory ramp is terminated, and then the algorithm explores the local gate voltage region by random walk, again while taking 2D scans. The score function evaluates these scans to determine the quality of the double-dot features the algorithm has found.}
    \label{fig: SEM}
\end{figure*}

\section{Methods}

\subsection{Device and rf readout} \label{sec: device}

We demonstrate our algorithm in a Ge/SiGe depletion mode quantum dot array operated as two double dot devices, labelled A and B, where five voltage gates shaped the electrostatic potential (Fig.~\ref{fig: SEM}(a)). The fabrication process of this device is similar to that described in \cite{jirovec_singlet-triplet_2021}. Ac and dc signals were applied to the plunger gates for each quantum dot, $V_2, V_4$ and $V_6$, via bias-tees. A constant voltage of \SI{2}{\volt} was set to the splitter gates $V_{\text{sl}}$ and $V_{\text{sr}}$ to isolate the triple dot array from a charge sensor, which was not used for this work. Measurements were performed at \SI{50}{\milli \kelvin}. 

One device ohmic was connected to an L-matching network for rf readout. This L-matching network was formed by a \SI{92}{\pico \farad} decoupling capacitor, a \SI{2.2}{\micro \henry} inductor, and the parasitic capacitance to the ground of the printed circuit board (PCB) and the device, $C_P$  (Fig.~\ref{fig: SEM}(a))\cite{ares_sensitive_2016}. These values were chosen to approximately match the impedance of the double quantum dot device \cite{vigneau_probing_2022}. The power of the rf drive at the device was fixed at \SI{-90}{dBm}. The rf signal's in-phase and out-of-phase components were acquired with homodyne demodulation.

The device measurements consisted of fast two-dimensional gate voltage scans. Plunger gates were swept in a raster pattern using an arbitrary waveform generator (Fig.~\ref{fig: SEM}(b)). These scans had an amplitude and resolution of 100 by \SI{100}{\milli \volt} and $100$ by $100$ pixels, respectively. An integration time of \SI{1}{\micro \second} per pixel was used, enabling an entire scan to be completed in \SI{10}{\milli \second}. It was necessary to precompensate the waveforms responsible for the raster pattern for the distortion introduced by the bias tees. See Appendix \ref{appendix: scans} for a complete description of how these scans were performed and precompensated. 

\subsection{Algorithm}

Our rf tuning algorithm is built from two components: a strategy to navigate through the device's gate voltage space and a score function to quantify the quality of any double dot features found. 

Previous strategies to navigate gate voltage spaces cannot be trivially applied to rf measurements, as matching networks are only sensitive over a small window of device impedances. These strategies used current measurement to search for the current hypersurface, which marks the transition between the device being `open' and `closed'~\cite{moon_machine_2020, severin_cross-architecture_2021}. Quantum dot features are found near this current hypersurface.

To circumvent this limitation, we formulate an rf hypervolume to navigate by. This volume is defined as where the rf matching network is sensitive to the changes in quantum device impedance. Outside the rf hypervolume, the matching network is not sensitive, and the signal in the demodulated in-phase and out-phase components falls below the noise floor. The current hypersurface and the rf hypervolume do not necessarily intersect, and we can control the position in the gate voltage space of the rf hypervolume by choice of the matching network components. With an appropriate choice of components, quantum dot features should be found near the rf hypervolume. In general, a range of occupations can be accessed depending on the rf circuit, including the few-carrier regime \cite{petersson_charge_2010}. In our case, these features correspond to carrier occupancies on the order of 10. 

Our algorithm navigates to the rf hypervolume using exploratory gate voltage ramps. These ramps slowly (at approximately \SI{1}{\volt \per \second}) move gate voltages towards \SI{0}{\volt} along a randomly-chosen radial direction in the space defined by voltage gates $V_1 - V_5$ for device A, or $V_3 - V_7$ for device B (Fig.~\ref{fig: SEM}(c)). The first of these ramps starts from a gate voltage coordinate on the largest sphere bounded by the gate voltage upper limits, which are 4V for all gates. This upper bound was chosen to span the full operational range of the gate electrodes constrained by leakage currents. Along the length of a ramp, the algorithm performs fast 2D gate voltage scans (described in \ref{sec: device}) every \SI{50}{\milli \volt}. The algorithm evaluates each of these scans using a classifier to determine whether the scan shows only noise. When the ramp crosses into the rf hypervolume, the classifier will cease to detect noise, and the ramp is terminated.

The algorithm then searches for double-dot features near the gate voltages where the rf hypervolume was found using a random walk. Double dot features are identified and quantified using a score function based on the discrete-time two-dimensional Fourier transform (described in \ref{sec: score}). The random walk is performed by perturbing a randomly-chosen gate voltage by an amount sampled from a Gaussian distribution with a standard deviation of \SI{50}{\milli \volt}.

The algorithm efficiently explores the gate voltage space by repeatedly undertaking these exploratory ramps. With each ramp, it has the possibility of finding new double-dot features which might score better than any dot features found previously. Over time, with repeated exploratory ramps, the algorithm explores an increasing fraction of gate voltage space; therefore, the score associated with the highest-scoring double-dot features should improve over the algorithm's run time. Concurrently, each ramp's termination gate voltage coordinate is used to update a Gaussian process, as in Ref. \cite{moon_machine_2020}. These observations of the hypervolume location inform the starting gate voltage coordinate of subsequent ramps, allowing them to start closer to the hypervolume and waste less time measuring regions of gate voltage space where only noise can be observed. Therefore, as the algorithm improves its model for the location of the hypervolume, it becomes more efficient at exploring the gate voltage space. 

The following sections describe the rf classifier and the score function used by the algorithm.

\subsubsection{The rf Classifier}\label{sec: rf_classifer}

 \begin{figure}
\centering
\includegraphics{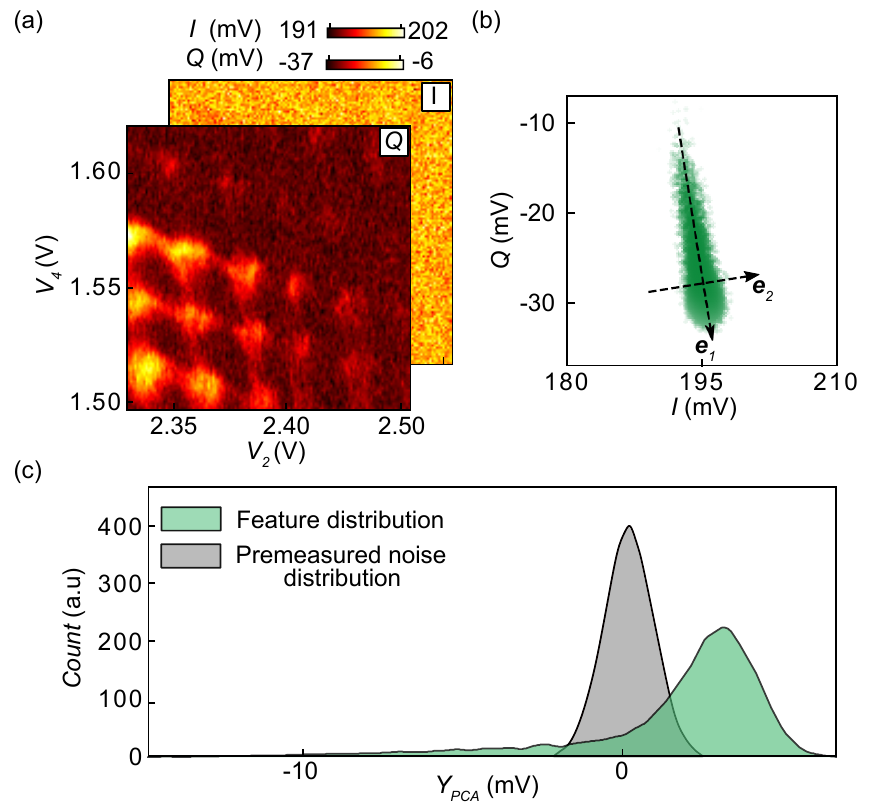}
\caption{(a) The demodulated in-phase $(I)$ and out-of-phase $(Q)$ quadratures of a 2D scan, showing double quantum dot features on the edge of the rf hypervolume. (b) A scatter graph of the demodulated $I$ and $Q$ quadratures of the 2D scans in (a). We observe two principal directions of variance. The information-bearing component can be found along $e_1$, while the less informative component is along  $e_2$. (c) A histogram of the $I$ and $Q$ data once projected onto $e_1$, shown in green. For comparison, we show in grey an histogram of pre-measured noise projected similarly. The noise has a Gaussian distribution, while the double dot features produce a distinctly different distribution, providing a means to distinguish them. \label{fig: FNC}}
\end{figure}

The role of our rf classifier is to evaluate whether one of our fast 2D scans shows noise or quantum dot features. If a scan at a given gate voltage coordinate shows quantum dot features, then that coordinate must lie within the rf hypervolume. Both in-phase ($I$) and out-of-phase ($Q$) components of the rf signal could contain useful information about these features (Fig.~\ref{fig: FNC}(a)). 
This can be regarded as a blind signal separation (BSS) problem since we wish to extract the information-bearing signal shared between the two quadratures. We use a popular method to deal with this type of problem known as principal component analysis (PCA). PCA considers the signal's covariance matrix to optimally find the direction in the $I-Q$ plane with the largest variance and projects the data onto it~\cite{tharwat_independent_2021}. Our algorithm uses PCA to optimally project a pair of 2D scans measured in both quadratures into a main one (Fig.~\ref{fig: FNC}(b)). This optimally-projected data is then classified using a hypothesis test to compare the pixel value distribution of the projected scan against a reference noise measurement. The noise is premeasured by taking 10 2D scans with all gate voltages set to their maximum value, far from the rf hypervolume. As a way to illustrate this test, a histogram of the PCA-projected pixel values of the scan in Fig.~\ref{fig: FNC}(a) is overlaid on the measured reference noise distribution (Fig.~\ref{fig: FNC}(c)). Quantum dot features can be observed in Fig.~\ref{fig: FNC}(a), and a long-tailed distribution captures this compared to the noise in Fig.~\ref{fig: FNC}(c). This 2D scan would thus be classified as containing features and terminate an exploratory gate voltage ramp.

The hypothesis test is a two-sample Kolmogorov-Smirnov test~\cite{hodges_significance_1958, scipy}. The null hypothesis is that the two distributions are identical; therefore, the 2D scan contains just noise. The alternative hypothesis is that the distributions differ, so the scan shows features instead.  Our critical p-value is $10^{-3}$, so p-values below this critical value lead to the rejection of the null hypothesis in favour of the alternative. This choice of p-value means that, on average, the null hypothesis is falsely rejected only once in every $10^3$ scans.

\subsubsection{Score Function}\label{sec: score}

The role of the score function is to identify scans showing double dot features by scoring them highly whilst scoring other features, such as those from single dots, poorly. Score functions used in the literature to evaluate current measurements have been based on neural networks ~\cite{kalantre_machine_2019, zwolak_autotuning_2020, ziegler_toward_2022}, Hough transforms \cite{lapointe-major_algorithm_2020,mills_computer-automated_2019}, custom fitting models \cite{baart_computer-automated_2016}, handcrafted heuristics \cite{moon_machine_2020}, and even ray-based classification \cite{zwolak_2021}. These score functions are too noise-sensitive for fast measurements or cannot generalise to complex signals. We designed a score function based on the discrete-time Fourier transform of the 2D scans. This score function is naturally phase-independent and thus can evaluate both in-phase and out-of-phase scans. Owing to the properties of the Fourier transform, it is also robust against noise, allowing us to benefit from the high bandwidth of rf measurements. See Appendix \ref{appendix: score} for a mathematical description of how the score function was evaluated and a discussion of the noise tolerance inherited from the Fourier transform. 

\begin{figure}
\centering
\includegraphics{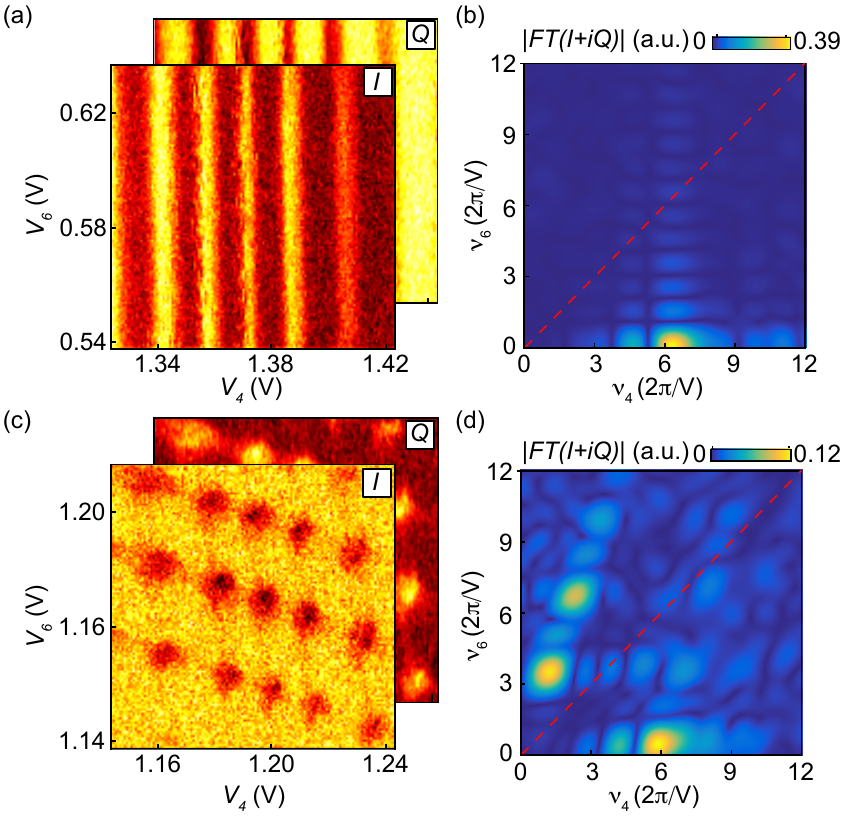}
\caption{(a-b) A 2D scan showing single dot features and corresponding time-discrete Fourier transform of the standardised $I$ and $Q$ data. The single dot features show one distinct periodicity, resulting in a single strong maximum in the Fourier transform (with fringes resulting from the finite window size).  (c-d) A 2D scan showing double dot features and the corresponding time-discrete Fourier transform of the standardised $I$ and $Q$ data. The double dot features show two distinct periodicities (and associated harmonics), resulting in two strong maxima in the Fourier transform on either side of the diagonal $\nu_x = \nu_y$ marked as a dashed red line.  \label{fig: score}}

\end{figure}

The discrete-time Fourier transform of the 2D scans quantifies any periodicities present in these scans. Each periodicity results in a prominent maximum in the Fourier transform, and the more periodic the features, the higher the value of these maxima. Our score function searches the Fourier transform of 2D scans for the two maxima resulting from double quantum dot features. We constrain the search for these maxima so that we only consider periodicities consistent with two quantum dots, each coupled enough to their respective plunger gates, showing a few charge transitions over a \SI{100}{\milli \volt} plunger gate sweep. In practice, if the plunger gates are $V_x$ and $V_y$, we search Fourier frequencies $\nu_x, \nu_y \in [0, 12]$ a.u. and only considering maxima on opposite sides of $\nu_x=\nu_y$. The score function then assigns the score as the strength of the weaker of the two maxima. A scan showing single dot features should result in just one maximum and other features in none, and thus low scores will be assigned to such scans. A comparison of both double dot and single dot features with respective Fourier transforms is displayed in Fig.~\ref{fig: score}(a-d). 

\section{Results}

To evaluate the performance of our algorithm both in devices A and B, we performed 24 runs for a fixed time of an hour. This resulted in the algorithm attempting to tune devices A and B 13 and 11 times, respectively, which produced $\sim 2\times 10 ^4$ scans in total, each with an associated score. To demonstrate the success of these runs, we need first to establish that the score function is a good metric for the identification of double quantum dot regimes and then show that the score increases as a function of tuning time and that it does so faster than more straightforward search strategies.

To quantify how correlated our score function was with experts' perception of double dot quality, we created three categories: non-double, imperfect double, and satisfactory double dot. Two human experts classified the scans according to such labels without knowledge of the associated score. Fig.~4(a) shows a stacked histogram of the scores, where each bar has been divided into colours based on the human experts' labels. We observe that non-double dot (satisfactory quantum dot) features reliably receive low (high) scores.

Based on Fig.~\ref{fig: score}(a), we define two score thresholds to be 0.04 and 0.08. These values are chosen to produce confusion matrices reflecting the score function's performance and have no bearing on the double quantum dot regimes that the algorithm finds. Tables \ref{table: A} and \ref{table: B} show the confusion matrices for devices A and B, respectively. Our score function produces very few false positives even when faced with an overwhelming number of scans in which no double quantum dot features are present whilst retaining the ability to separate satisfactory from imperfect double quantum dots.

\begin{table}[]
\caption{Confusion matrix for device A.}
\label{table: A}
\begin{tabular}{|cl|ccc|}
\hline
\multicolumn{2}{|c|}{\multirow{2}{*}{Device A (\%)}}   & \multicolumn{3}{c|}{Expert Classification}                                                                                                                                              \\ \cline{3-5} 
\multicolumn{2}{|c|}{}             & \begin{tabular}[c]{@{}c@{}}Non-double \\ dot\end{tabular} & \begin{tabular}[c]{@{}c@{}}Imperfect\\double dot\end{tabular} & \begin{tabular}[c]{@{}c@{}}Satisfactory\\double dot\end{tabular} \\ \hline
\multicolumn{2}{|c|}{score $< 0.04$}              & 96.96                                                     & 0.93                                                           & 0.00                                                       \\
\multicolumn{2}{|c|}{$0.04 \leq $ score $< 0.08$} & 0.19                                                      & 1.43                                                           & 0.06                                                       \\
\multicolumn{2}{|c|}{score $\geq 0.08$}           & 0.00                                                      & 0.11                                                           & 0.30                                                       \\ \hline
\end{tabular}
\end{table}

\begin{table}[]
\caption{Confusion matrix for device B.}
\label{table: B}
\begin{tabular}{|cl|ccc|}
\hline
\multicolumn{2}{|c|}{\multirow{2}{*}{Device B (\%)}}   & \multicolumn{3}{c|}{Expert Classification}                                                                                                                                              \\ \cline{3-5} 
\multicolumn{2}{|c|}{}             & \begin{tabular}[c]{@{}c@{}}Non-double \\ dot\end{tabular} & \begin{tabular}[c]{@{}c@{}}Imperfect\\double dot\end{tabular} & \begin{tabular}[c]{@{}c@{}}Satisfactory \\double dot\end{tabular} \\ \hline
\multicolumn{2}{|c|}{score $< 0.04$}              & 96.93                                                     & 0.89                                                           & 0.00                                                       \\
\multicolumn{2}{|c|}{$0.04 \leq $ score $< 0.08$} & 0.30                                                      & 1.41                                                           & 0.09                                                       \\
\multicolumn{2}{|c|}{score $\geq 0.08 $}           & 0.01                                                      & 0.10                                                           & 0.27                                                       \\ \hline
\end{tabular}
\end{table}

\begin{figure}
\centering
\includegraphics{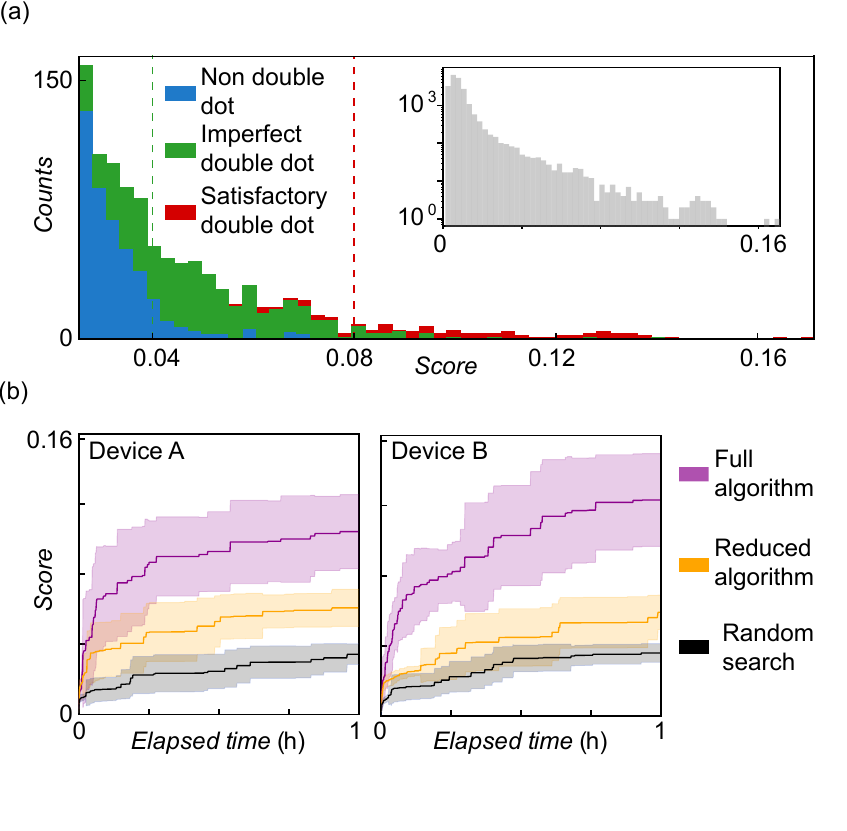}
\caption{(a) A histogram of the double dot feature scores plotted on a linear scale, where the colour coding of each bin indicates relative proportions of the scans judged by human labelling to be non-double (blue), imperfect double (green) and satisfactory double dot (red). The red double dot region lies above the blue non-double dot region, whilst the green region corresponding to satisfactory double dots lies above both the orange and the blue. For clarity, we plotted only scores above $0.25$ on the linear scale due to the overwhelming number of scores below this threshold; the inset shows a histogram of all the scores on a log scale. (b-c)  The highest score found for increasing tuning times for algorithms which randomly search (grey), a reduced version of the algorithm (orange) and the full algorithm (pink) in devices A and B. To account for the statistical variation between runs, the curves presented are the mean of many runs. Our algorithm explores the gate voltage space more efficiently than a random search or random rays; therefore, its best score improves fastest.    \label{fig:5}}

\end{figure}

We now show the improvement of the score with time. Over its run time, the algorithm explores a greater fraction of the rf hypervolume and discovers more dot features. Each newly-discovered dot feature has a chance of scoring higher than anything found before; therefore, on average, the score associated with the highest-scoring features should improve over time. The purple curves in Fig.~\ref{fig:5}(b-c) show the evolution of the best score found by the algorithm as a function of run time for devices A and B, respectively. For a baseline comparison, we ran a reduced version of our algorithm, which uses the exploratory ramps but did not benefit from any of the Gaussian processes, the rf classifier (which terminated the gate voltage ramps close to the rf hypervolume), nor the subsequent random walk, shown in orange. We also ran a random search algorithm that randomly chose a coordinate in gate voltage space, set the gate voltages, and then performed a scan, shown in black.

We find that our algorithm outperforms these more simplistic versions. This is likely because our algorithm spends a more significant fraction of its time exploring regions that have the potential to correspond to double dot regimes. 

Overall, the full algorithm found and identified 16.5 and 16.8 scans an hour on average, corresponding to double dots (satisfactory and non-ideal) in devices A and B, respectively. If we restrict our attention to satisfactory double dots, the rates are 2.4 and 2.7 scans an hour. These tuning times are the fastest recorded in laterally-defined quantum dots \cite{moon_machine_2020, severin_cross-architecture_2021}. See Appendix \ref{appendix: textbook} for plots of satisfactory double dots tuned by the algorithm in each hour-long run.

\section{Conclusion}

In conclusion, we have demonstrated the first algorithm for tuning double dots exclusively utilising rf measurements. Such a result is an essential first step toward the scalability of quantum devices. Our algorithm reliably tuned satisfactory double quantum dots dot in approximately 15 minutes, the fastest recorded time in laterally-defined quantum dots. In creating this algorithm, we developed a robust method for distinguishing noise from rf features and proposed a score function based on the Fourier transform with significant noise tolerance. 

Faster device tuning times could be achieved using a more informative Gaussian prior for the Gaussian process, particularly one which encodes knowledge of the gate architecture. Such a prior could use an electrostatic model of the device~\cite{craig_bridging_2021}. Furthermore, an additional Gaussian process and Bayesian optimisation techniques could be used to choose the ramp directions.

To probe the charge states of large quantum dot arrays, the fast 2D scans could be generalised to radial rays, as performed slowly in \cite{chatterjee_autonomous_2021, zwolak_2021}. Along with frequency multiplexing \cite{hornibrook_frequency_2014}, our algorithm could allow for scalable tuning and, thus, fully-automatic tuning of multiple small quantum dot arrays.

\begin{acknowledgments}
   We thank Nicholas Sim for providing help with the experiment and David Craig for his feedback on the manuscript. This work was supported by the Royal Society (URF-R1-191150), the EPSRC National Quantum Technology Hub in Networked Quantum Information Technology (EP/M013243/1), Quantum Technology Capital (EP/N014995/1), EPSRC Platform Grant (EP/R029229/1), the European Research Council (Grant agreement 948932), the Scientific Service Units of IST Austria through resources provided by the nanofabrication facility, the FWF-P 30207 and FWF-I 05060 projects, and grant number FQXi-IAF19-01 from the Foundational Questions Institute Fund, a donor advised fund of Silicon Valley Community Foundation.
\end{acknowledgments}

\FloatBarrier

\appendix

\section{High Resolution Scans}\label{appendix: scans}

In this section, we describe how fast two-dimensional scans were performed. 
For this task, we used a Tektronix AWG (model 5014c) to feed a slow (red) and a fast (blue) sawtooth waveforms to the device gates $V_2$, $V_4$ and $V_6$, as shown in Fig.~\ref{fig: SEM}(b), such that their combination would produce a raster scan in the gate voltage plane defined by $V_2$ and $ V_4$ ($V_4$ and $V_6$) for device A (B), centred around the gates dc-voltage. Since the amplitude of the ramp determines the scan size, this is ultimately limited by the amount of attenuation in the cryostat lines. We chose our scans to be $100\times100$ \SI{}{\milli \volt} in size.


Measurements were performed by rf reflectometry, using a Zurich lock-in amplifier (UHFLI) to feed a frequency tone of 137.5 MHz to the L-matching network connected to
the device ohmic and demodulate the reflected component of the signal. The resulting $I$ and $Q$ quadratures were finally recorded using a fast digitiser card (Alazar tech ATS9440) using an AWG marker to trigger the card acquisition such that it would be synchronised with the voltage ramps. We used an integration time per pixel of $\tau_{\text{int}} = $ \SI{1}{\micro \second} and a resolution of 100 by 100 pixels, allowing us to perform a full 2D scan within \SI{10}{\milli \second}.  

The exact shape of the sawtooth waveform needed to be adjusted to avoid artefacts introduced by high-pass filters in the setup. The action of a first-order high-pass filter with a time constant $\tau$ on a signal $s(t)$ is $s_{\text{filtered}}(t) =\mathcal{F}^{-1}[\mathcal{F}[s(t)](\omega)\cdot H_{HP}(\omega)]$ where $H_{HP}(\omega)=i\omega\tau / (1+i\omega\tau)$, and $\mathcal{F}[\cdot]$ denotes the time to frequency domain Fourier transform. To correct for this effect, we pre-compensated the waveforms according to $s_{\text{compensated}}(t)=\mathcal{F}^{-1}[\mathcal{F}[s(t)](\omega)\cdot H_{HP}^{-1}(\omega)]$, such that the final shape reaching the device gates would be exactly $s(t)$. Precompensating the desired waveform for this filter requires only the knowledge of the filter's time constant.
At cryogenic temperatures, we measured the time constants to be $1.043,1.051,1.025$ \SI{}{\milli \second} for gates $V_2$, $V_4$ and $V_6$, respectively. Given that the slow sawtooth waveform period was \SI{10}{\milli \second}, precompensation was necessary.

 
\section{Projection of $I$ and $Q$ data with principal component analysis} \label{appendix: pca}

When performing the fast 2D scans, we measured both the in-phase and out-of-phase quadratures. In general, both components could carry information.
At the price of recording both quadratures, PCA extracts the information-carrying signal from the noise~\cite{comon_2010} and eliminates the need of phase adjustment. 
Let us consider a 2D scan with in-phase data, $\vec{I}$, and the out-of-phase data, $\vec{Q}$, where each index in these vectors corresponds to a pixel in the 2D scan. To perform blind signal separation via principal component analysis on this data, we obtained the eigenvector, $\vec{e}_1$, corresponding to the largest eigenvalue of the covariance matrix,
\begin{equation}
	\left[\begin{array}{cc}
		\cov(\vec{I},\vec{I}) & \cov(\vec{I},\vec{Q})\\
		\cov(\vec{Q},\vec{I}) & \cov(\vec{Q},\vec{Q})
	\end{array}\right]. 
\end{equation}
We then obtained the extract the information bearing-signal, $\vec{Y}_{\text{PCA}}$, by projecting the in-phase and out-of-phase data onto this eigenvector, such that 
\begin{equation}
	\vec{Y}_{\:\text{PCA}} = \vec{e}_1^{\;\intercal} \mathbf{X} \quad \text{where} \quad \mathbf{X} = \begin{bmatrix}
		(\vec{I} - \bar{I}) \rightarrow \\
		(\vec{Q} - \bar{Q})\rightarrow 
	\end{bmatrix},
	\end{equation}
where $\bar{I}$ and $\bar{Q}$ denote the mean of $\vec{I}$ and $\vec{Q}$ respectively. 

\section{The discrete time Fourier transform score function}\label{appendix: score}

This section describes how to evaluate the DTFT score function on the in-phase, $\vec{I}$, and out-of-phase, $\vec{Q}$, data produced by a 2D scan of resolution $N\times M$ pixels. Firstly, we combine the in-phase and out-of-phase components into one complex dataset, to obtain $\vec{Z} := \vec{I} + i\vec{Q} \in\mathbb{C}^{N \times M}$. Then, we standardise this complex data, $\vec{Z}$ according to $\vec{\mathcal{Z}} := (\vec{Z}-\bar{Z}) / {\sigma_Z}$, where $\bar{Z} := \bar{I} + i \bar{Q}$ and $\sigma^2_Z := \vari[I] + \vari[Q]$. The reason for standardising the data is so that the score function reflects the quality of the double dot features and not how well-matched they are to the L-matching circuit. Next, we compute the discrete-time Fourier transform of $\vec{\mathcal{Z}} $ according to 
\begin{equation}
		\text{DTFT}[\vec{\mathcal{Z}}]_{\delta \gamma}=\frac{1}{NM}\sum_{a =1}^{N}\sum_{b =1}^{M}e^{ -2\pi i(\nu^{x}_{\delta}a+\nu^{y}_{\gamma}b)} \mathcal{Z}_{ab}, \label{eq:DTFT}
	\end{equation}
where $\vec{\nu}^x$ and $\vec{\nu}^{y}$ are a high-resolution linspace of frequencies between 0 and 12, to obtain $\text{DTFT}[\vec{\mathcal{Z}}]\in\mathbb{C}^{100\times100}$. From here the score, $s$, is computed according to 
\begin{equation}
	s := \min \left(\max |\text{DTFT}[\vec{\mathcal{Z}}]|_{\delta < \gamma},\max|\text{DTFT}[\vec{\mathcal{Z}}]|_{\delta > \gamma}\right).
\end{equation}
To paraphrase this formula, the score function finds the strongest periodicity presented above and below the line $\nu_x = \nu_y$ and then assigns the score to be the weaker of the two. In this manner, the score function favours double dots where each dot is coupled most strongly to its plunger gate. If there is just a single dot present, then there could be strong periodicity and even perhaps harmonics on one side of the line $\nu_x = \nu_y$. However, single-dot features cannot score highly without a dot coupled to the opposite plunger, as the two periodicities necessary for a high score would not be present. 
 
As discussed in the following section, this score function inherits considerable noise tolerance from the Fourier transform, such that the noise in the score function is suppressed compared to the noise in 2D scans by a factor of the square root of the number of pixels, $NM$. In our work, we used $N$ and $M=100$, meaning that the noise in the elements of the DTFT was suppressed by a factor of 100. 

\subsection{The discrete-time Fourier transform's noise characteristics}

If $\vec{Z} \in \mathbb{C}^{N\times M}$ is a complex dataset of $NM$ elements and $\vec{N} \in \mathbb{C}^{N\times M}$ is the corresponding gaussian noise, where each element is complex-normally distributed such that $N_{ij}\sim\mathcal{\mathcal{C}N}(0,\sigma_N)$ (meaning that the real and imaginary parts of elements of $\vec{N}$ are independently normally distributed with a mean of zero and variance of $\sigma_N ^2 /2$). Then 
 	\begin{equation} \label{eq: DTFT_noise}
	\text{DTFT}[\vec{Z}+\vec{N}]_{\delta \gamma}\sim\mathcal{\mathcal{C}N}\left(\text{DTFT}[\vec{Z}]_{\delta \gamma},\frac{\sigma_N}{\sqrt{NM}}\right),
\end{equation}
where $\text{DTFT}[\cdot]$ is the discrete-time Fourier transform. Therefore, the noise is suppressed by a factor of $1 / \sqrt{NM}$ in the discrete-time Fourier transform. This can be shown by using the additive nature of the discrete-time Fourier transform. We can write $\text{DTFT}[\vec{Z}+\vec{N}]_{\delta \gamma}=\text{DTFT}[\vec{Z}]_{\delta \gamma}+\text{DTFT}[\vec{N}]_{\delta \gamma}$, where the time-discrete Fourier transform of the noise $\vec{N}$ is
\begin{equation}
	\text{DTFT}[\vec{N}]_{\delta \gamma}= \frac{1}{NM}\sum_{a =1}^{N}\sum_{b =1}^{M} e^{-2\pi i(\nu^{x}_{\delta}a+\nu^y_{\gamma}b)} N_{ab}.
\end{equation}
As the complex-normal distribution is circularly-symmetric the multiplication by the complex phase factor, $\exp(-2\pi i(\nu^{x}_{\delta }a+\nu^{y}_{\gamma}b))$, leaves distribution unchanged, thus $\exp(-2\pi i(\nu^{x}_{\delta }a+\nu^{y}_{\gamma}b)) N_{ab} \sim\mathcal{CN}(0,\sigma_N)$, which means that $\text{DTFT}[\vec{N}]_{\delta \gamma}$ can be regarded as the mean of $NM$ samples from a complex-normal distribution. The central limit theorem predicts that this mean will be distributed according to $\text{DTFT}[\vec{N}]_{\delta \gamma}\sim\mathcal{\mathcal{C}N}\left(0,\sigma_N / \sqrt{NM} \right)$. As there is no uncertainty in the Fourier transform of the noiseless signal, $\vec{N}]_{\delta \gamma}$ the sum of the Fourier transform of the signal and noise will be distributed according to equation \ref{eq: DTFT_noise}.

\section{Outcomes from the hour-long tuning sessions}\label{appendix: textbook}

\begin{figure*}
	\centering
    \includegraphics{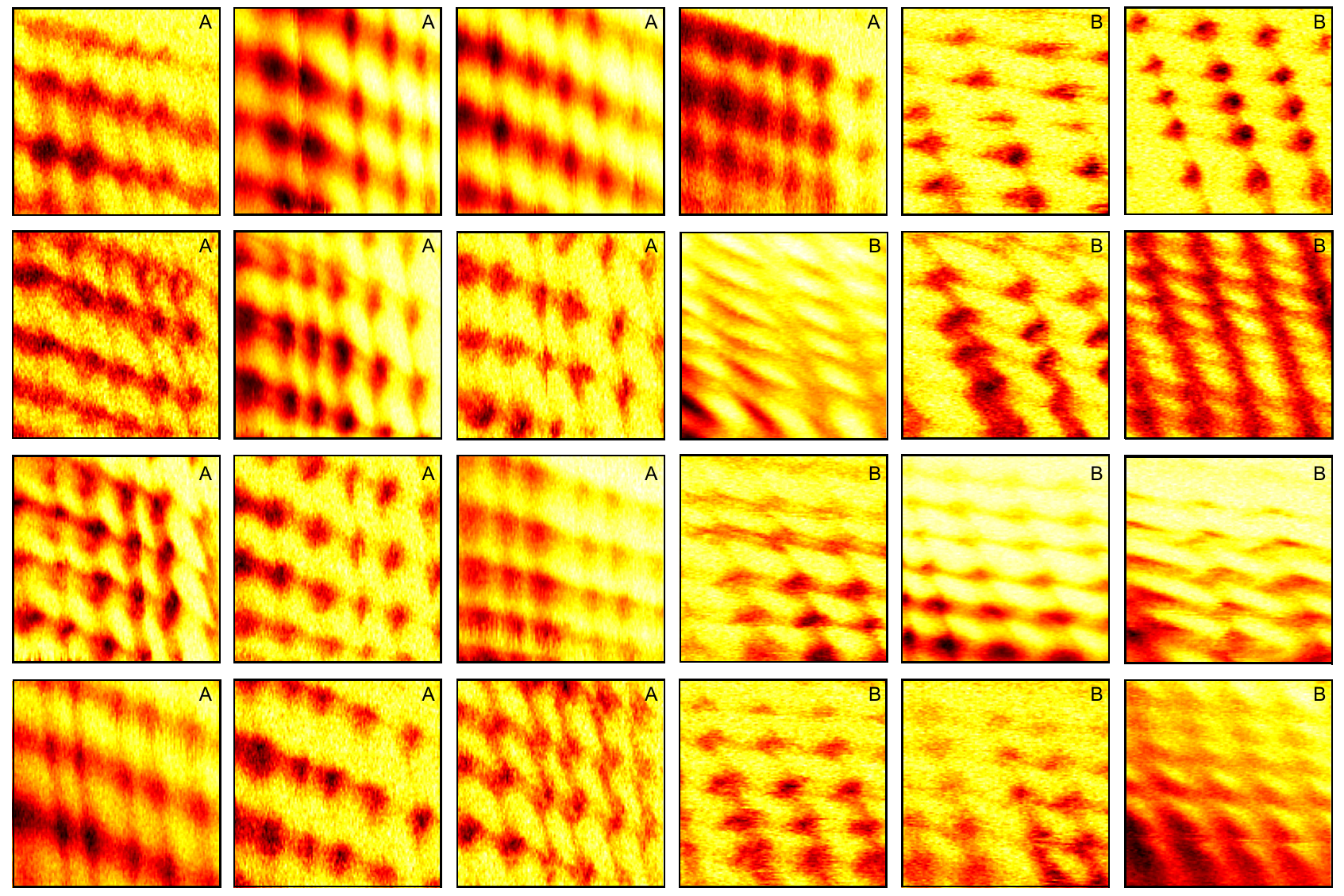}
	\caption{The highest-scored double dot features found by the algorithm by the end of each of the hour-long tuning sessions performed on devices A or B, labelled in the top corner. For scans taken in device A (B), the horizontal and vertical axis are gates $V_2$ ($V_4$) and $V_4$ ($V_6)$ respectively, both of these gate voltages are swept over a \SI{100}{\milli \volt} range. The pixel values correspond to the PCA projection described in appendix \ref{appendix: pca}. \label{appendix: outcomes} 
}
\end{figure*}

The highest scoring of the double dot features found by the algorithm in each hour-long tuning session (Fig.~\ref{appendix: outcomes}). 

\FloatBarrier

\end{document}